\documentclass[aps,pre,reprint,superscriptaddress]{revtex4-1}

\pdfoutput=1

\usepackage{graphicx}
\usepackage{amssymb}
\usepackage{amsmath}
\usepackage{xcolor}
\usepackage{pdfpages}
\usepackage{pgffor}
\usepackage{CJK}
\usepackage{multirow}
\usepackage{gensymb}

\makeatletter
\AtBeginDocument{\let\LS@rot\@undefined}
\makeatother

\begin{document}

\title{On the Nature of Freezing/Melting Water in Ionic Polysulfones} %Title of paper

\author{Britannia Vondrasek}
\affiliation{Macromolecules Innovation Institute, Virginia Polytechnic Institute and State University, Blacksburg, Virginia 24061, USA}
\author{Chengyuan Wen}
\affiliation{Macromolecules Innovation Institute, Virginia Polytechnic Institute and State University, Blacksburg, Virginia 24061, USA}
\affiliation{Department of Physics and Center for Soft Matter and Biological Physics, Virginia Polytechnic Institute and State University, Blacksburg, Virginia 24061, USA}
\author{Shengfeng Cheng}
\email{chengsf@vt.edu}
\affiliation{Macromolecules Innovation Institute, Virginia Polytechnic Institute and State University, Blacksburg, Virginia 24061, USA}
\affiliation{Department of Physics and Center for Soft Matter and Biological Physics, Virginia Polytechnic Institute and State University, Blacksburg, Virginia 24061, USA}
\affiliation{Department of Mechanical Engineering, Virginia Polytechnic Institute and State University, Blacksburg, Virginia 24061, USA}
\author{Judy S. Riffle}
\affiliation{Macromolecules Innovation Institute, Virginia Polytechnic Institute and State University, Blacksburg, Virginia 24061, USA}
\affiliation{Department of Chemistry, Virginia Polytechnic Institute and State University, Blacksburg, Virginia 24061, USA}
\author{John J. Lesko}
\affiliation{Macromolecules Innovation Institute, Virginia Polytechnic Institute and State University, Blacksburg, Virginia 24061, USA}
\affiliation{Department of Mechanical Engineering, Virginia Polytechnic Institute and State University, Blacksburg, Virginia 24061, USA}
%\email{jlesko@vt.edu}

\date{\today}

\begin{abstract}
We investigate the behavior of hydrated sulfonated polysulfones over a range of ion contents through differential scanning calorimetry (DSC), Fourier transform infrared spectroscopy (FTIR), and molecular dynamics (MD) simulations. Experimental evidence shows that at comparable ion contents, the spacing between the ionic groups along the polymer backbone can significantly impact the amount of melting water present in the polymer. When we only consider water molecules that can hydrogen bond to four neighboring water molecules as the melting water, the MD simulation results are found to agree with the experimental data. The states of water measured by DSC can therefore be described as ``aggregated'' (or bulk-like) for the melting component, and ``isolated'' for the nonmelting part. Using this physical picture, a polymer with more aggregated ions has a higher content of melting water, while a polymer at the same ion content but with more dispersed ions has a lower content of melting water. Therefore, ions should be well dispersed to minimize the amount of bulk-like water in ionic polymer membranes.
\end{abstract}

\maketitle

\section{Introduction}
\label{Intro}

\par Ionic polysulfone-based membranes have potential applications in a wide range of fields including fuel cells,\cite{Harrison2005} ionic actuators,\cite{Tang2014} electrolysis of water,\cite{daryaei2017-1} reverse osmosis,\cite{daryaei2017-2} and electrodialysis.\cite{choudhury2019} In these applications, the hydration behavior of the polymer is a critical parameter determining the functionality of the membrane. More specifically, ion transport, electrical resistance, thermal, mechanical, and other bulk properties of a solid ionic polymer have been found to depend not only on the amount of water present in the membrane, but also on the specific way in which the absorbed water interacts with the polymer chains and fixed ions.\cite{tran2019, woo_2003} However, our current understanding of the hydration mechanism of ionic polysulfones and its relationship to the bulk properties of the resulting membranes is limited, which hampers our ability to use them across a broader range of operating conditions.\cite{kocherbitov2010}

\par Differential scanning calorimetry (DSC) is a common method of probing the hydration characteristics of polymers. In this method, a small amount of hydrated polymer sample is packed in a sealed pan and the difference in the heat flux needed to increase the temperature of the sample and an empty reference pan is measured. To analyze hydration, the sample is cooled to below the freezing point of water. The sample is then heated at a set heating rate in order for any frozen water to melt. This procedure results in a freezing exotherm and/or a melting endotherm for water in the sample. From these features, the enthalpy of water freezing or melting can be calculated by taking the area under the exotherm or endotherm. It is more common to analyze the melting enthalpy data from an endotherm as the freezing exotherm sometimes is absent or splits into multiple peaks.\cite{karlsson2002, wu2010} 

\par DSC studies indicate that water within a hydrated ionic polymer can be classified into three general categories:\cite{luck1980, ping2001, roy2017} 
\begin{enumerate}
	\item Non-freezing or nonmelting water that does not contribute to the melting endotherm or freezing exotherm when probed by DSC. It is described as ``bound" water and hypothesized to consist of water molecules that interact most strongly with the polymer.
	\item Intermediate water that freezes and melts, but not at 0\degree C. Therefore it has a melting enthalpy that differs from that of bulk water. This type of water is sometimes described as ``loosely-bound" water.
	\item Bulk-like or ``free" water that melts and freezes at 0\degree C. This category of water is hypothesized to have minimal to no interaction with the polymer.
\end{enumerate}
The above classification scheme is a framework for interpreting DSC data collected for hydrated ionic polymers. However, there is not uniform support for the hypothesis that non-freezing water comprises the water molecules that are in direct contact with the ionic groups or hydrogen bonded to the hydrophilic sites on the polymer.\cite{ping2001, luck1980, karlsson2002} While experimental evidence does reveal features such as longer residence times and restricted rotational dynamics of water molecules at the interface with ionic solutes, there is little evidence to support the claim that such water molecules are immobilized to the extent that they cannot participate in ice formation even at low temperatures.\cite{zhang_molecular_2017, braun_rotational_2016, marque2010, dogonadze_chemical_1985, Charkhesht2018, Charkhesht2019} Furthermore, while some have observed that the appearance of bulk-like water within a polymer substantially impacts its physical properties, the specific relationship between the amount of melting water and the bulk properties of the hydrated polymer remains poorly understood.\cite{pastorczak2014, kocherbitov2016}

\par Fourier transform infrared spectroscopy (FTIR) is another experimental method that has been used to probe the behavior of water in hydrated polymers.\cite{ping2001, thouvenin_2002, lasagabaster_2006, smedley_measuring_2015} Other vibrational spectroscopy methods such as Raman, microwave, and terahertz spectroscopy also appear to be increasingly utilized because of advancements in quantum chemical calculation of molecular vibrational frequencies and the application of vibrational multivariate curve resolution.\cite{ben-amotz_2019, verma_2018, tainter_2013} Impedance-based methods such as broad band and microwave dielectric spectroscopy can complement vibrational absorbance spectra and enhance our understanding of the dynamics of polar molecules such as water.\cite{luo_engineering_2019,luo_functional_2019} Despite current work in this area, the subpeaks obtained from deconvolution of the OH stretch band cannot be ascribed an exact physical meaning. In general, FTIR vibrational signatures at lower wavenumbers correspond to structures that are heavier or more constrained, while vibrational signatures at higher wavenumbers are from lighter or less constrained structures. Therefore, the shape of the OH stretching region is related to the variety of interactions that the water hydrogens experience and can be taken as an indicator of the extent of hydrogen bonding in the system. This general idea of more constrained compared to less constrained motions of water molecules forms the basis for the four-state model of the OH stretching band revealed by FTIR, which we employ in our analysis.\cite{tainter_2013, skinner_2008_2, thouvenin_2002, lasagabaster_2006} Additional insight into the state of water sorbed in ionic polymers can be obtained by using HDO instead of H$_2$O as the OD stretching has a single, well-defined vibrational band with a peak that depends on the local environment HDO molecules experience.\cite{smedley_measuring_2015}

\par The combination of experimental measurements and molecular dynamics (MD) simulations can help us more thoroughly understand the behavior of water in hydrated polysulfones at a molecular level. A number of MD simulations of hydrated aromatic polymers have been reported previously.\cite{venkatnathan2007,vishnyakov2008, marque2010, pozuelo2006, merinov2013, huo2019, bahlakeh2013, BAHLAKEH2012} Some earlier atomistic investigations of hydrated Nafion and sulfonated polystyrene were conducted by Venkatnathan \textit{et al.} \cite{venkatnathan2007} and Vishnyakov \textit{et al.}.\cite{vishnyakov2008} Hydration of the nonionic polysulfone backbone was investigated by Marque and co-workers.\cite{marque2010} Pozuelo \textit{et al.} \cite{pozuelo2006} and Merinov \textit{et al.} \cite{merinov2013} both present MD simulations of ionic polysulfones in the hydrated state. In MD simulations of ionic polymers, water content is typically expressed as $\lambda$, which is the number of water molecules per cation-anion pair. Huo \textit{et al.} performed MD simulations to study the effect of water uptake on the hydrogen-bonding network and its relation to OH$^-$ conduction in an imidazolium-based polymer.\cite{huo2019} Their results showed that when $\lambda$ is below 10, almost all the absorbed water molecules are within the primary hydration structures of the ions. For $\lambda > 10$, the fraction of water not within the primary hydration shells starts to grow and the polymer begins to swell.\cite{huo2019} Bahlakeh \textit{et al.} studied sulfonated polysulfone blended with sulfonated poly(ether ether ketone) using MD simulations.\cite{BAHLAKEH2012, bahlakeh2013} They simulated a range of hydration levels of $\lambda$=3, 6, 9, 12, and 15 and found evidence of larger-scale water clusters for $\lambda \gtrsim 9$.\cite{bahlakeh2013}

\par Previous studies have already revealed a variety of interrelated physical phenomena that can be associated with a transition in the behavior of water in ionic polymers, such as the formation of water clusters, the percolation of hydrogen-bonding networks, and the filling of higher-order hydration shells around the fixed and mobile ions.\cite{tripathy_molecular_2017, bahlakeh2013, roy2006} In all of these phenomena, the distribution of ions within the polymer plays a crucial role. However, until recently there was no reliable way to isolate the spatial distribution of ions as a variable. By combining advanced ionic polymer synthesis, atomistic modeling, and an understanding of phase separation in ionic polymers, we can now look beyond the average, bulk ion density and consider the impact of local ion distributions on the behavior of absorbed water and the properties of ionic polymers. Even in desalination membranes based on uncharged polymers, recent studies showed that the distribution of functional groups plays an important role in controlling salt transport.\cite{luo_engineering_2019, luo_functional_2019}

\par In this paper, DSC and FTIR experiments are combined with MD simulations to study the physical nature of the transition from a state without to a state with freezable water in ionic polysulfones. We investigate the effect of ion spacing along the polymer backbone on the amount of melting water, and then use simulation results to gain a deeper understanding of the molecular mechanisms underlying the experimental data. This enhanced understanding of the nature of freezing/melting water will contribute to future work exploring the link between the hydration behavior and bulk physical properties, such as ion exclusion and mechanical performance, of ionic polysulfone membranes.

\section{Materials and Methods}

\subsection{Materials}
\label{Materials}

\begin{figure*}[htb]
\center
\includegraphics[width=\textwidth]{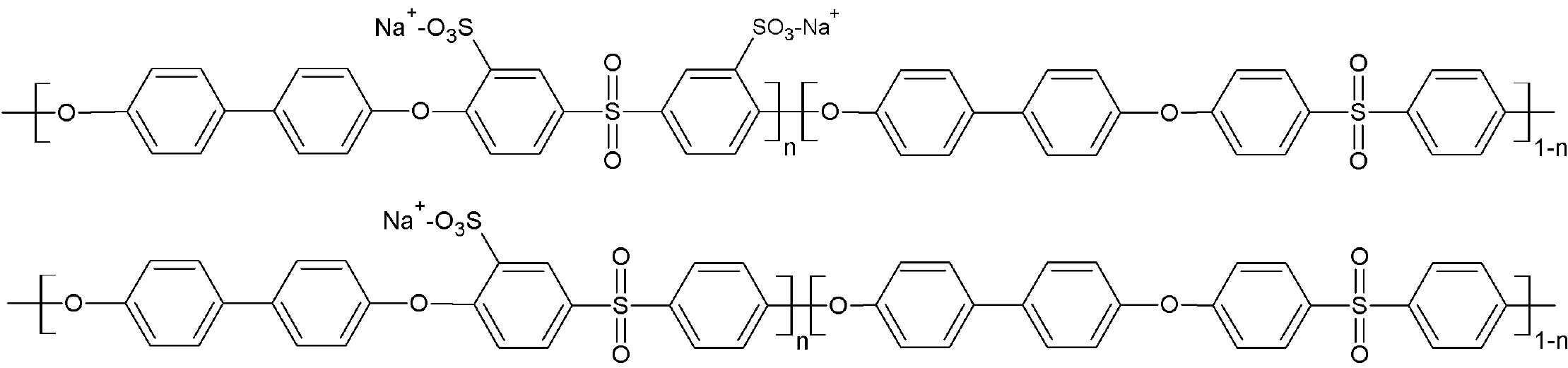}
\caption{Chemical structures of the polysulfones used in the experiments: dBPS - disulfonated (top) and mBPS - monosulfonated (bottom).}
\label{fig:structure_exp}
\end{figure*}

\par We investigate a series of sulfonated polysulfones that are of particular interest for water purification applications, including reverse osmosis and electrodialysis.\cite{daryaei2017-2, choudhury2019} Presulfonated sulfone monomers are used to synthesize the polymers, which are subsequently exchanged to their sodium counterion form as described elsewhere.\cite{daryaei2017} The chemical structures of the polymers are shown in Fig.~\ref{fig:structure_exp}. Their properties are summarized in Table~\ref{TB1}. The only difference between the monosulfonated and disulfonated polymers is how the sulfonate ions are distributed along the polymer backbone. In the monosulfonated polymer, there is only one ion per sulfonated monomer while in the disulfonated polymer, a sulfonated monomer has two sulfonate ions. We use mPBS$x$ to denote a monosulfonated polysulfone and dBPS$x$ for a disulfonated one, where $x$\% is the fraction of repeat units that are sulfonated. Ion content is therefore $x$\% for mBPS$x$ and $2x$\% for dBPS$x$.

\par These polymers are well suited for fundamental studies of polymer-water interactions for a number of reasons. The high content of aromatic rings on the backbone makes polysulfones much stiffer than other ionic polymers commonly reported in the literature, such as Nafion.\cite{daryaei2017-1} The prevalence of aromatic rings results in a high glass transition temperature ($T_g$). For a dry polysulfone with a similar structure to those discussed in this paper, $T_g$ is above 220 \degree C. Even when hydrated, the value of $T_g$ for all the sulfonated polysulfones studied here remains above 120 \degree C as noted in Table \ref{TB1}. One benefit of $T_g$ being high is that the primary thermal transition of the polymer occurs in a completely different temperature range than the phase change of water in the polymer, which makes the analysis of DSC data on melting water more straightforward.\cite{muller-plathe_1998}

\begin{table*}[htb]
\centering
\begin{tabular}{|c|c|c|c|c|c|c|c|}
\hline
        & Molecular & Ion     & Mass Water &     &                & Hydrated   & Hydrated  \\
Polymer & Weight    & Content & Uptake     & IEC$^b$ & $\lambda_\text{eq}$ & $T_g$ & Density   \\
        & ($M_\text{w}$, kDa)     & (Mol \%)    & (Wt \%)       &   (meq/g)  &  & (\degree C) & (g/cm$^3$) \\ \hline
BPS0 & 120         &  0       & 2.6        & 0     & 0      & --   &  1.33    \\ \hline
mBPS50 & 127       & 50       & $14 \pm 1$   & 1.16  & 8.4    & 157  &  1.34   \\ \hline
mBPS61 & 93        & 61       & $21  \pm 3$  & 1.36  & 10.6   & 164  &  1.34    \\ \hline
mBPS79 & 45        & 79       & $31  \pm 4$  & 1.70  & 13.7   & 137  &  1.33   \\ \hline
dBPS22 & 97        & 44       & $13  \pm 3$  & 0.99  & 7.5    & 177  &  --  \\ \hline
dBPS27 & 144       & 54       & $18  \pm 2$  & 1.18  & 8.3    & 169  &  --  \\ \hline
dBPS32 & 51        & 64       & $25  \pm 3$  & 1.37  & 9.9    & 164  &  -- \\ \hline
dBPS33 & --$^a$    & 66       & --$^a$     & --$^a$& --$^a$ & --$^a$ & 1.35\\ \hline
dBPS40 & 107       & 80       & $32 \pm 1$   & 1.65  & 12.4   & 141   & 1.33\\ \hline
\end{tabular}
\caption{Properties of the sulfonated polysulfones in their sodium salt forms. $^a$Polymer labeled dBPS33 is a different batch of polymer that is assumed to be similar to dBPS32 in terms of properties. $^b$Reported IEC is measured experimentally by titration.}
\label{TB1}
\end{table*}

\par In addition to high $T_g$, the bond angle of the sulfone linkages contributes to preventing polysulfone chains from crystallizing, resulting in completely amorphous polymers. Therefore, the polymer-water interactions are not complicated by the possible variations in crystallinity in a particular sample.\cite{ping2001} Also because of the stiff and heavy backbone, polysulfones exhibit relatively low water uptake at a given ion content, which makes them ideal for investigating water that is most strongly associated with the polymers. Finally, there is no experimental or theoretical evidence of microscale ionic aggregation or phase separation in sulfonated polysulfones.\cite{sundell2014, vondrasek2021} As a result, analysis of the absorbed water is not likely to be impacted by phase inhomogeneities.

\subsection{Water Uptake and $\lambda_\text{eq}$}
\label{wuandlambda}

\par Ionic polysulfone samples are prepared as described elsewhere.\cite{daryaei2017-1, vondrasek2021} To measure the equilibrium water uptake ($W_\text{eq}$), three samples, each approximately 3 cm by 3 cm, are cut from a hydrated film. The samples are blotted with a laboratory wipe to remove surface water, and then quickly weighed. The samples are dried at 150\degree C under vacuum for 48 hours and then allowed to cool to room temperature while still under vacuum. The dry samples are quickly removed from the oven and weighed again. The samples are immersed in deionized (DI) water for at least 36 hours for re-hydration before the blotting and weighing processes are repeated. The equilibrium water uptake, $W_\text{eq}$, for each sample is calculated as
\begin{equation}
    W_\text{eq} = \frac{M_\text{wet}-M_\text{dry}}{M_\text{dry}}~,
    \label{eq:wu}
\end{equation}
where $M_\text{wet}$ and $M_\text{dry}$ are the masses of the wet and dry polymers, respectively. The water uptake and loss values for all the polymers being studied agree within the measurement errors. In other words, the water uptake is fully regained after drying and there is no significant hysteresis effect for these polymers. The water uptake results from the three samples are then averaged to obtain the final value of $W_\text{eq}$ for each polymer.

\par The ion exchange capacity (IEC) of a polymer membrane is defined as the number of sulfonate ions per gram of the hydrated polymer, expressed in meq/g. In this work we present IECs measured experimentally by titration. The equilibrium water uptake ($\lambda_\text{eq}$), defined as the number of water molecules per sulfonate-sodium ion pair, is computed as
\begin{equation}
    \lambda_\text{eq}=\frac{W_\text{eq}\times 1000}{\text{IEC}\times m_{\text{H}_{2}\text{O}}}~,
    \label{eq:lambda}
\end{equation}
where $m_{\text{H}_{2}\text{O}}$ is the molar mass of water (18.015 g/mol). It is important to note that $\lambda_\text{eq}$ represents the number of water molecules required to create the entire hydration structure of the fixed ion, the counterion, and the polymer repeat unit associated with a single ion pair. The value of $\lambda_\text{eq}$ therefore depends on the type of counterion present. All data presented in this work are for the $\text{Na}^{+}$ counterion form.

\par Table \ref{TB1} shows that $\lambda_\text{eq}$ increases with an increasing ion content. For solid polymers, the ions provide the primary driving force for hydration, and the stiffness of the polymer backbone is the primary factor opposing hydration. Therefore, as more ions are incorporated into a polymer, it can absorb more water, which in turn plasticizes the polymer. The plasticized polymer is more prone to swelling, which helps enhance the equilibrium water uptake. As a result, $\lambda_{eq}$ gets larger when the ion content becomes higher.

\subsection{DSC Measurement}
\label{DSC}

\par For DSC analysis, high-volume pans with a fluoropolymer O-ring from TA Instruments are used. They are first cleaned according to the recommended protocol.\cite{sloughnodate} The pan, O-ring, and lid are weighed prior to adding a polymer. The polymer film is stored in DI water prior to testing. Small pieces of the hydrated polymer film are cut and packed into the high-volume pan. The pans are filled with as much polymer as possible to minimize the formation of an air/water vapor mixture inside the pan during the test. The open pan is dried under vacuum overnight at 150\degree C, allowed to cool under vacuum for at least 8 hours, and then quickly weighed to determine the mass of the polymer in the pan. The required volume of water is calculated as
\begin{equation}
    V_{\text{H}_2\text{O}} = \frac{M_\text{dry} W_\text{c}}{\rho_{\text{H}_2\text{O}}}~,
\end{equation}
where $M_\text{dry}$ is the dry mass of the sample, $W_\text{c}$ is the intended water content expressed as a fraction, and $\rho_{\text{H}_2\text{O}}$ is the density of water at room temperature. DI water of volume $V_{\text{H}_2\text{O}}$ is then added to the pan using an adjustable micropipette and the pan is closed. The sealed pans are allowed to equilibrate at room temperature for at least 24 hours before testing.

\par To obtain DSC data, a sealed pan is first cooled at -5 \degree C/min to -85 \degree C and equilibrated. The pan is then heated to 150 \degree C, equilibrated, and held at that temperature for 5 minutes primarily as a relaxation step to remove any residual stress that might exist in the polymer. Subsequently, the pan is cooled again at -5 \degree C/min to -85 \degree C and equilibrated. A final temperature ramp to 250 \degree C at 5\degree C/min is conducted in order to obtain both water-melting data and $T_g$ for the hydrated polymers. The pans have no change in mass over the course of the experiment and therefore can be considered as closed systems.

\par Since the relaxation step is conducted above the boiling point of water, some water must be sorbed by the polymer when it is cooled down. Therefore, the water melting endotherm is sensitive to the cooling rate of the sample. At cooling rates higher than 5 \degree C/min, water does not have sufficient time to diffuse into the polymer as it cools, and therefore all polymers display much larger melting endotherms and their magnitudes vary widely with the cooling rate. At cooling rates lower than 5\degree C/min, the magnitude of the endotherms does not change substantially as the cooling rate is varied. Based on this, we conclude that a cooling rate of 5 \degree C/min is suitable for these polymers as it is slow enough to allow for their complete rehydration as they are cooled.

\par The number of melting water molecules per fixed sulfonate ion is calculated using the equation: 
\begin{equation}
    N_\text{m}=\frac{\lambda \times \Delta H_\text{m}}{W_\text{eq} \times \Delta H_{\text{H}_2\text{O}}(T_\text{m})}~,
    \label{eq:N_m}
\end{equation}
where $\Delta H_\text{m}$ is the melting enthalpy of water within the polymer and $\Delta H_{\text{H}_2\text{O}}(T_\text{m})$ is the standard melting enthalpy of water as a function of its melting temperature, $T_\text{m}$. Since $T_\text{m}$ changes between samples, $H_{\text{H}_2\text{O}}(T_\text{m})$ is taken as a function of $T_\text{m}$ using the equation, 
\begin{equation}
H_{\text{H}_2\text{O}}(T_\text{m})=0.0037\times T_\text{m}^2+0.0902\times T_\text{m}-0.4647~.
\end{equation}
This equation was obtained from data presented by Ruscic.\cite{Ruscic_2013} In this work, we consider a single value of $T_\text{m}$, which is defined as the temperature at the maximum point in the water melting endotherm.

\begin{figure*}[htb]
\centering
\includegraphics[width=\textwidth]{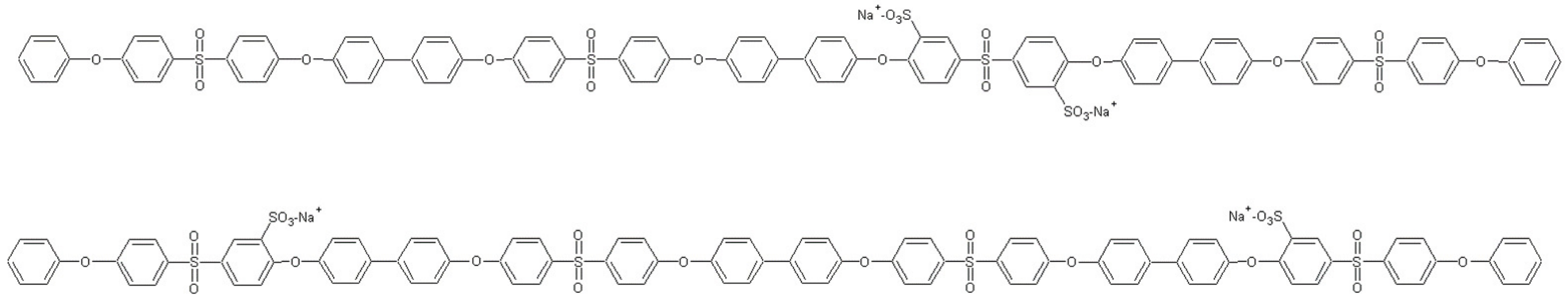}
\caption{Structures of the disulfonated (top) and monosulfonated (bottom) tetramers used for MD simulations.} 
\label{fig:structure_MD}
\end{figure*}

\subsection {Attenuated Total Reflection Fourier Transform Infrared (ATR-FTIR) Measurement} \label{ATR_FTIR}

For ATR-FTIR spectroscopy measurements, saturated film samples are stored in DI water until just before testing. The film is removed from the DI water and its surface is lightly dried with a lab wipe before being placed on the ATR crystal. A Thermo Fisher Nicolet iS50 FTIR spectrometer is used to obtain the spectra in ATR mode. Each measurement is the average of 64 scans and a new baseline is collected prior to each run. ATR correction is applied to the spectra, but automatic baseline correction is not applied because the water libration band at low-wavenumbers causes the baseline correction to be inconsistent. The multipeak fitting package in Igor Pro software is used to fit the water OH stretch band between 1430 and 1370 cm$^{-1}$ using Gaussian peak types.

\subsection{MD Modeling}

\par The chemical structures of the disulfonated and monosulfonated oligomers simulated are shown in Fig.~\ref{fig:structure_MD}. MD simulations are performed as described elsewhere.\cite{vondrasek2021} Both disulfonated and monosulfonated polymers are simulated at an ion content of 50\% (consistent with the dBPS25 and mBPS50 polymers). Each MD system consists of 64 tetramers (comprising 128 fixed sulfonate ions and 256 sulfone groups), 128 sodium counterions, and $128\lambda$ water molecules. We also present the simulation results for mBPS75, which is constructed in the same way as mBPS50, but with an extra sulfonate ion on each tetramer. Table~\ref{TB1} shows the experimental values of the equilibrium water uptake, $\lambda_\text{eq}$. In the simulations, $\lambda$ is varied from 3 to 14 in order to span the entire range of water contents probed experimentally.

\section{Results}

\subsection{DSC Results}
\label{DSC_VIC}

\begin{figure}[htb]
    \centering
    \includegraphics[width=0.45\textwidth]{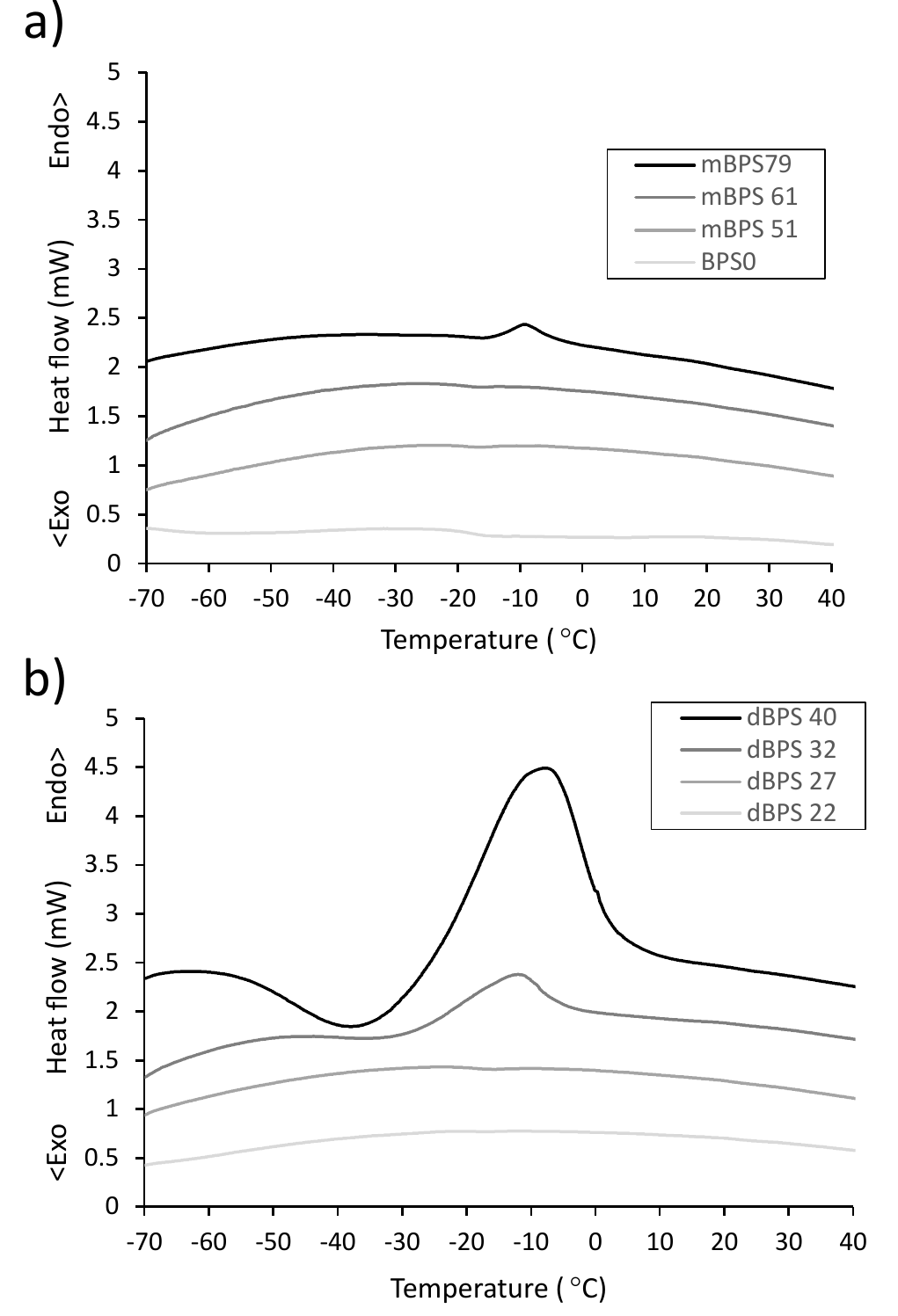}
    \caption{DSC curves showing melting behavior for the (a) monosulfonated and (b) disulfonated polysulfones of varying ion contents at equilibrium water uptake. For comparison, the data for the nonsulfonated polysulfone, BPS0, is also included in (a).}
    \label{fig:DSC}
\end{figure}

\par
\begin{figure}[htb]
    \centering
    \includegraphics[width=0.45\textwidth]{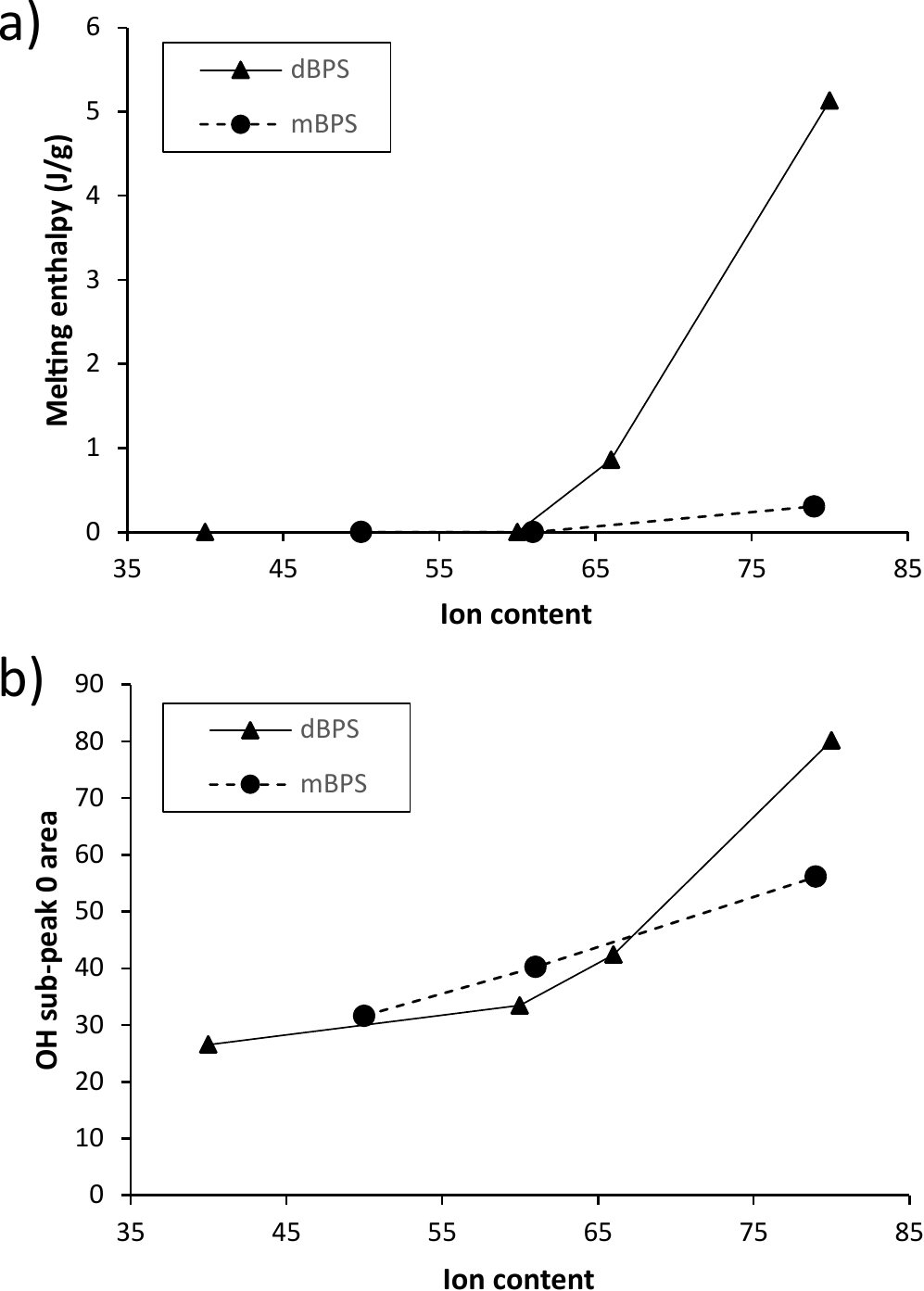}
    \caption{(a) Melting enthalpy, computed from the DSC curves, of water in the sulfonated polysulfones with varying ion contents. (b) Area of the subpeak 0, reflecting the low-wavenumber contribution, from the FTIR OH stretching band of the sulfonated polysulfones as a function of ion content. The lines are guides to the eye.}
    \label{fig:DSCFTIRcomp}
\end{figure}

 The DSC curves for the sulfonated polysulfones of varying ion contents at their equilibrium water uptake are shown in Fig.~\ref{fig:DSC}. The small transition near -20 \degree C is visible in the run with an empty pan and therefore attributed to the fluoropolymer O-ring of the high-volume DSC pans. For the polymers with low ion contents (mBPS50, mBPS61, dBPS22, and dBPS27), the equilibrium water uptake is low and no melting endotherm appears. On the other hand, for dBPS32, dBPS40, and mBPS79, the equilibrium water uptake increases and a melting endotherm is observed. The resulting average melting enthalpy is shown as a function of ion content in Fig.~\ref{fig:DSCFTIRcomp}(a).

\par The melting enthalpy, which is proportional to the amount of melting water in the sample, shown in Fig.~\ref{fig:DSCFTIRcomp}(a) reveals a clear contrast between the monosulfonated and disulfonated polymers. As discussed in Sec.~\ref{Materials}, the only difference between the two polymers at the same ion content is the spacing of ions along the backbone. If the nonmelting water only consists of water molecules that are associated with the ions, then the amount of melting water in a sulfonated polysulfone should be predominantly determined by its ion content and not affected by the ion spacing. However, the data in Fig.~\ref{fig:DSCFTIRcomp}(a) indicate that the ion spacing has a significant impact on the amount of melting water in a polymer. The melting enthalpy is essentially 0 for the monosulfonated polymers with low ion contents and only slightly deviates from 0 for the mBPS79 polymer. However, for the disulfonated polymers, the melting enthalpy increases significantly when their ion contents become high enough ($\gtrsim 60\%$), such as the dBPS33 and dBPS40 polymers with ion contents at about 66\% and 80\%, respectively. The trends revealed by the DSC measurements are further supported by the FTIR data shown in Fig.~\ref{fig:DSCFTIRcomp}(b), which is discussed in the next section.

\subsection {FTIR Results}
\label{FTIR_VIC}

\par
\begin{figure}[htb]
    \centering
    \includegraphics[width=0.45\textwidth]{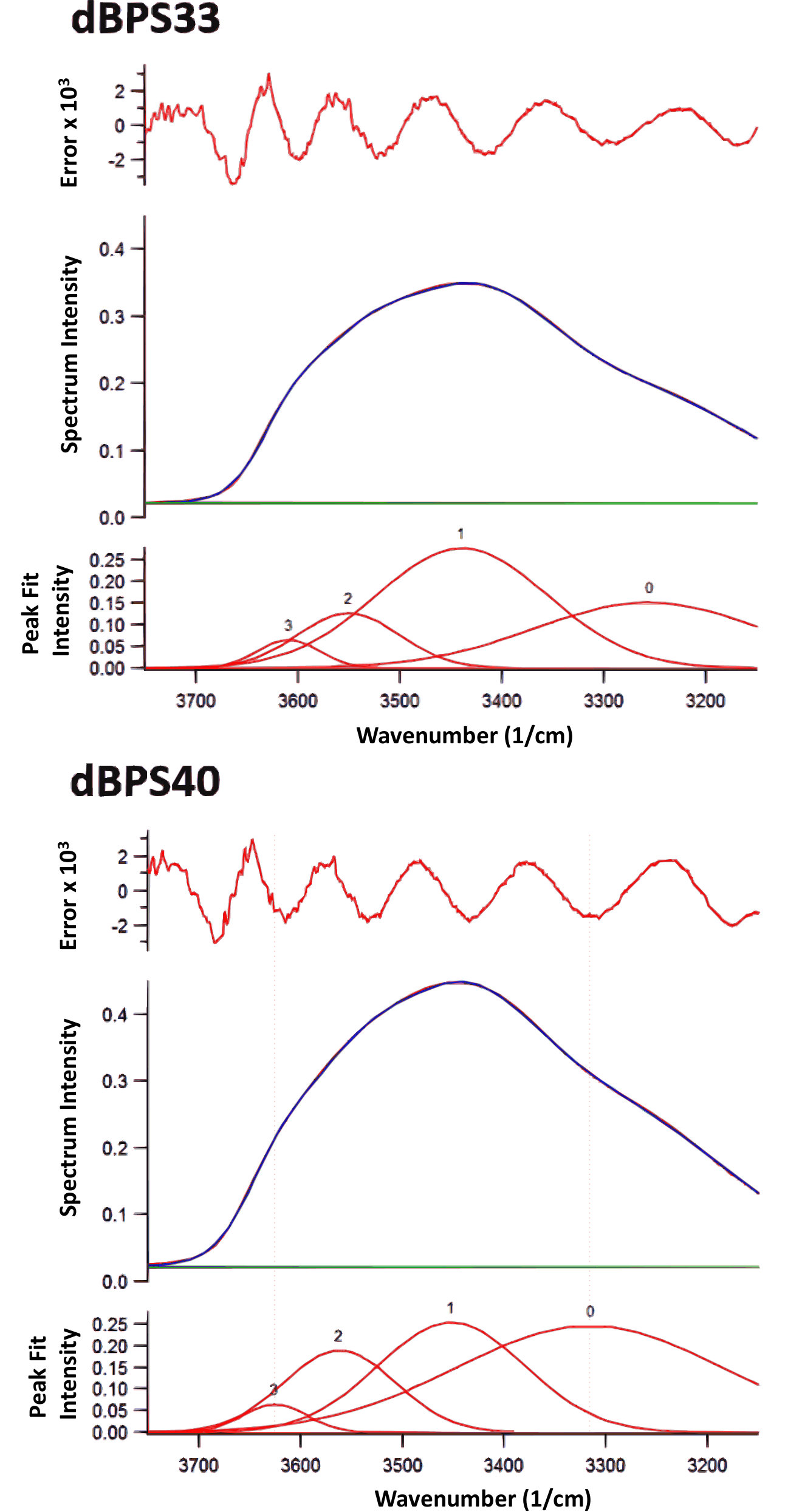}
    \caption{Deconvolution of the FTIR OH stretching band for the dBPS33 (top) and dBPS40 (bottom) polymers. The relative size of the subpeak 0 shows a significant increase for dBPS40.}
    \label{fig:OHdeconv}
\end{figure}

 The OH stretching band from the FTIR measurements for two disulfonated polymers, dBPS33 and dBPS40, and their respective deconvolutions are shown in Fig.~\ref{fig:OHdeconv}. The FTIR OH stretching region is deconvoluted into four subpeaks according to the four-state model.\cite{skinner_2008_2} Several other deconvolution techniques were also attempted, but the four-state model was the only one that made it possible to fit the signature of polymers with both low and high ion contents using similarly located peaks. Using this model, the subpeaks 0 and 1 represent more hindered OH stretching vibrations as would occur in bulk water. The subpeaks 2 and 3 represent less hindered OH stretching vibrations exhibited by water molecules with incomplete hydrogen bonding, such as those near a hydrophobic surface.\cite{auer_ir_2008} The data show that for dBPS33, the dominant mode is the subpeak 1 while for dBPS40, it is the subpeak 0.

\par With the understanding that melting water is more bulk-like, we compute the area under the subpeak 0 for all the sulfonated polymers, which reflects the contribution from bond vibrations in the low-wavenumber region. This area is expected to be directly related to the amount of melting water in a hydrated polymer. The data are plotted in Fig.~\ref{fig:DSCFTIRcomp}(b). For the monosulfonated polysulfones, this area grows with the ion content gradually in the entire range of ion contents being probed. However, for the disulfonated polysulfones, the area under the subpeak 0 exhibits a sharp increase at ion contents higher than about 60\%, specifically for dBPS33 and dBPS40. These trends, including the difference between the monosulfonated and disulfonated cases, are consistent with those revealed by the DSC data in Fig.~\ref{fig:DSCFTIRcomp}(a) and therefore corroborate the finding that there is a transition in the amount of melting water in the disulfonated polysulfones at high ion contents. This transition, on the other hand, is not observed for the monosulfonated polymers, at least in the range of ion contents studied here.

\par Both DSC and FTIR measurements reveal a transition in the amount of melting water in the sulfonated polysulfones when the ion content (i.e., the level of sulfonation) is increased. The physical origin of this transition is unclear. Furthermore, within the probed range of ion contents, the transition is much more apparent for the disulfonated polymers. This difference in behavior between the monosulfonated and disulfonated polysulfones is related to the difference in the ion spacing along the polysulfone backbone but the molecular-scale understanding of the connection is still lacking.\cite{vondrasek2021} The interpretation of the experimental data should be carefully considered. The shape of the OH stretch region from FTIR measurements may be sensitive to the hydrogen-bonding environment of water molecules and their interactions with ions. As a result, the area underneath the subpeak 0 represents the combined contributions from the water molecules with four hydrogen bonds as well as other water molecules with hindered stretching vibrations. Therefore, it is unclear whether the increase in the peak area of the FTIR spectrum reflects a change in the ion-association state of water in the polymer or an increase in the number of water molecules with a bulk-like hydrogen-bonding environment. To answer these questions, we conduct detailed MD simulations of polysulfone oligomers, which are constructed to simulate the hydration behavior of the polysulfones studied in the experiments.

\subsection{Comparison to MD Simulations}
\label{MD}

\par In the experimental measurements discussed above, all the sulfonated polymers are at equilibrium water uptake, i.e., $\lambda = \lambda_\text{eq}$, which depends on their ion content. In this sense, the results plotted in Fig.~\ref{fig:DSCFTIRcomp}(a) can also be viewed as the variation of the amount of melting water with respect to the total water uptake in the polymers. However, this perspective is complicated by the fact that the polymers used to collect the data presented in Fig.~\ref{fig:DSCFTIRcomp}(a) have different ion contents. From a computational point of view, it is much more straightforward to simulate a sulfonated polysulfone at a fixed ion content and vary the amount of water in the simulation.\cite{vondrasek2021} To obtain experimental data that can be directly compared to such simulations, we focus on two disulfonated polymers, dBPS33 and dBPS40, which show a distinct amount of melting water at their equilibrium water uptake. We hydrate these polymers at various water contents and then run DSC measurements as described in Section \ref{DSC}. The DSC curves and the resulting melting enthalpy, similar to those shown in Fig.~\ref{fig:DSCFTIRcomp}(a), confirm that for both polymers, no melting water is present at low water contents while at high water uptakes, the amount of melting water starts to increase almost exponentially. The transition appears to occur at approximately $\lambda \simeq 6 \sim 7$.

\begin{figure}[htb]
    \centering
    \includegraphics[width=0.45\textwidth]{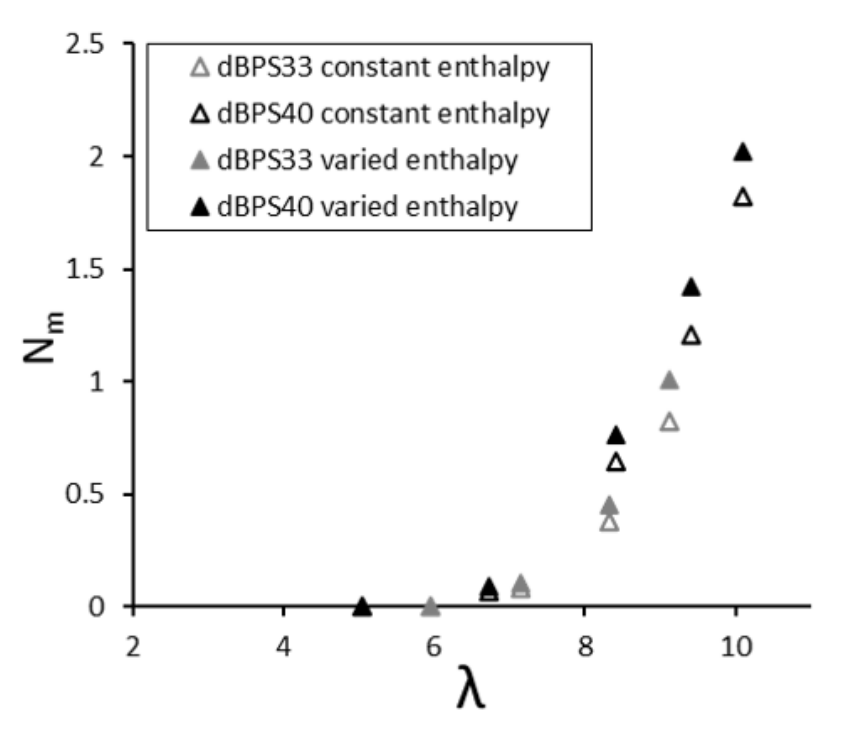}
    \caption{Number of melting water molecules ($N_\text{m}$) per ion pair as a function of water uptake ($\lambda$) for dBPS33 (light gray) and dBPS40 (dark gray). The data are converted from the melting enthalpy data from the DSC measurements, using either the value of melting enthalpy at 0 \degree C (open symbols) or a variable value that depends on the measured melting temperature (filled symbols).}
    \label{fig:contwater}
\end{figure}

\par To facilitate comparison with the MD results, we convert the melting enthalpy to the number of melting water molecules ($N_\text{m}$) per anion-cation pair and plot it against the water content expressed as $\lambda$ (i.e., the number of water molecules per ion pair). The detail of this conversion is expressed in Eq.~\ref{eq:N_m}. In the literature, this conversion is frequently done using the constant melting enthalpy of water at 0 \degree C.\cite{yangnodate} However, since the measured melting temperature of water in the sulfonated polysulfones may deviate from 0 \degree C (e.g., see the DSC curves in Fig.~\ref{fig:DSC}), we also perform this conversion using the melting enthalpy of water at the measured melting temperature, which corresponds to the location of the maximum in the DSC curve. In Fig.~\ref{fig:contwater}, we include results from both conversion methods, using either the constant melting enthalpy of water (at 0 \degree C) or a variable melting enthalpy. The two results are similar because when the measured melting enthalpy increases (i.e., as the amount of melting water increases), the melting temperature approaches 0\degree C, which causes the measured melting enthalpy of water to approach the constant value at 0 \degree C. Below we use the number of melting water molecules converted from the DSC curve using a variable melting enthalpy.

\par The magnitude of the $y$-axis in Fig.~\ref{fig:contwater} indicates that there is very little melting water in any of these polymers. Specifically, the most hydrated polymer that we have investigated (dBPS40 at equilibrium hydration) has only around two melting water molecules per fixed ionic group. Given the small amount of melting water in these polymers, virtually all sorbed water molecules should be considered as interacting with the polymer in some way. From this perspective, we do not observe truly ``free" or bulk water in the polysulfones being investigated. Instead, all water molecules exist on a continuum in terms of their interaction strength with the polymer.

\par As discussed previously, there are two main hypotheses for the physical origin of melting water molecules:
\begin{enumerate}
	\item[] Hypothesis 1:  The melting water consists of water molecules that are not in the primary hydration shell of the ions.
	\item[] Hypothesis 2:  The melting water molecules are those that exhibit a bulk-like hydrogen-bonding network.
\end{enumerate}
It is challenging to test these two hypotheses experimentally. However, MD simulations can directly resolve the distribution of water around ions and allow the hydrogen bonds between water molecules to be analyzed.\cite{vondrasek2021} Therefore, we can test these hypotheses by directly counting water molecules in MD simulations through analyzing either hydration shells or hydrogen-bond networks and comparing the simulation results to the experimental data.

\par
\begin{figure}[htb]
    \centering
    \includegraphics[width=0.45\textwidth]{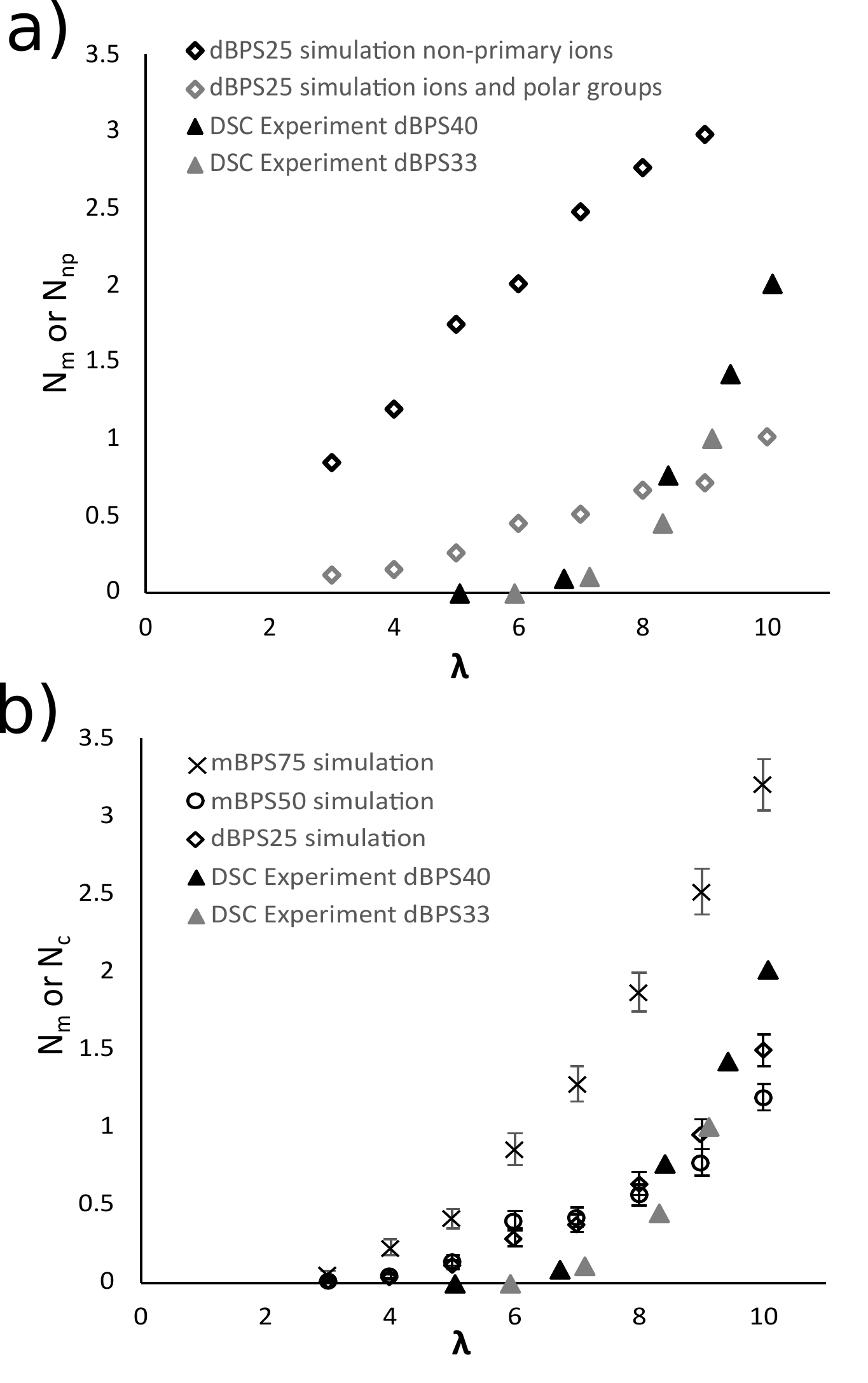}
    \caption{Comparison between the experimentally measured amount of melting water ($N_\text{m}$, light gray triangles for dBPS33 and dark gray triangles for dBPS40) and (a) the amount of ``nonprimary'' water molecules ($N_\text{np}$, light gray diamonds for water molecules outside the primary hydration shells of ions and polar groups and dark gray diamonds for water molecules outside the primary hydration shells of ions only) and (b) the amount of ``crystallizable'' water molecules ($N_\text{c}$, diamonds for dBPS25, circles for mBPS50, and crosses for mBPS75) from MD simulations as a function of water uptake ($\lambda$). The simulation data are averaged over a series of configurations and the error bars reflect the standard deviations.}
    \label{fig:compmelting}
\end{figure}

 MD simulations are conducted on the dBPS25, mBPS50, and mBPS75 polymers. The details of these simulations are described elsewhere.\cite{vondrasek2021} The only difference between mBPS50 and mBPS75 is that for the latter, each tetramer comprises three monosulfonated sulfone monomers rather than two. For all the polymers, a range of water uptakes is probed by changing the number of water molecules added to the system, with $\lambda$ varying from 3 to 14. To test the first hypothesis, we compute the number of water molecules that are outside of the primary hydration shells of $\text{Na}^+$ and $\text{SO}_3^-$ ions in MD simulations. Considering that water molecules are also attracted by the polar groups, we also count the water molecules that are outside of the primary hydration shells of all ions and polar groups, including the $\text{SO}_2$ groups and the O atoms in the ether linkages. Since these water molecules are not in a primary hydration shell, we refer to them as ``nonprimary", and their number, normalized by the number of sulfonate groups in the system, is designated $N_\text{np}$. The calculated values for $N_\text{np}$ from MD simulations are plotted against $\lambda$, and compared with the experimental data on $N_\text{m}$ in Fig.~\ref{fig:compmelting}(a). The two data sets show distinct trends. When calculated using only the ions, $N_\text{np}$ is much too high. When the polar groups are also included, the magnitude of $N_\text{np}$ is reduced, but the simulation data still clearly deviate from the experimental results. More importantly, $N_\text{np}$ computed on the basis of hydration shells varies roughly linearly with $\lambda$ and no clear transition can be identified as $\lambda$ is increased. This linear trend can be understood by noting that when all ions and polar groups are sufficiently hydrated with their primary hydration shells filled, additional water molecules added to a system will become ``nonprimary''. In contrast, the experimental data on $N_\text{m}$ has a clearly nonlinear dependence on $\lambda$ and increases significantly as the water content is increased beyond a certain threshold (i.e., for $\lambda \gtrsim 6$).

\par Next, we consider the second hypothesis that the melting water molecules are those with a bulk-like hydrogen-bonding environment. In order to identify hydrogen bonds in the simulations, we define two water molecules as hydrogen-bonded when the separation of their oxygen atoms is less than 3.4~\AA, which is approximately the inter-oxygen distance of neighboring water molecules in bulk water. The same distance is also used to quantify water clustering.\cite{tripathy_molecular_2017, zhang_molecular_2017, bahlakeh2013} For each water molecule, we identify all water molecules to which it is hydrogen-bonded. By this metric, water molecules with four (or more which is rare) hydrogen bonds are considered to have a bulk-like hydrogen-bonded network (i.e., two donor bonds and two acceptor bonds), regardless of their proximity to an ionic or polar group. We refer to these water molecules as ``crystallizable" water because they fulfill one of the primary criteria for forming the tetrahedral crystalline structure of ice. The number of such water molecules is denoted $N_\text{c}$ and the results from MD simulations are plotted as a function of $\lambda$ in Fig.~\ref{fig:compmelting}(b), together with the experimental data on $N_\text{m}$.

\par As observed in Fig.~\ref{fig:compmelting}(b), when the water uptake is increased, the amount of ``crystallizable'' water ($N_\text{c}$) for the dBPS25 polymer grows with a similar nonlinear trend as the amount of melting water ($N_\text{m}$) extracted from the DSC data on melting enthalpy, though some small quantitative difference can be observed. Considering that only tetramers are used in MD simulations, the agreement between simulation and experiment is qualitatively acceptable and indicates that it is reasonable to regard water molecules with a bulk-like hydrogen-bonding environment as melting water. From a physical perspective, this conclusion is sensible as melting and freezing are cooperative processes reflecting the collective motion of many water molecules. On the other hand, just one water molecule or a small water cluster outside the primary hydration shells of the ions and polar groups in a polymer matrix, and thus not strongly bound to them, will not contribute to melting or freezing probed by DSC measurements as their modes of motion can significantly deviate from those in bulk water.

\par The simulation results on the number of crystallizable water molecules, $N_\text{c}$, for the mBPS75 polymer are also included in Fig.~\ref{fig:compmelting}(b). A similar nonlinear dependence on $\lambda$ is observed but the value of $N_\text{c}$ is larger than that for the dBPS25 polymer at the same water uptake. Note that $\lambda$, $N_\text{c}$, and $N_\text{m}$ are all normalized by the number of ion pairs. The amount of melting water derived from the DSC curves is also slightly higher for dBPS40 than for dBPS33 at the same water uptake, indicating that the amount of melting water per ion-pair increases as the ion content is increased at a fixed $\lambda$. Also note that mBPS75 has a higher ion content at 75\% than dBPS25 (at 50\%). The trend revealed by MD simulations that mBPS75 has more melting water per ion-pair than dBPS25 at the same water uptake is therefore qualitatively consistent with the experimental results.

\par The simulation results in Fig.~\ref{fig:compmelting}(b) show that at a given water content, mBPS50 and dBPS25 contain almost the same amount of crystallizable water based on the hydrogen-bond criterion. This result is consistent with the DSC and FTIR data in Fig.~\ref{fig:DSCFTIRcomp}. For the polymers modeled here with an ion content of 50\%, there is no appreciable difference in the behavior of hydration water as the ion content is low. The difference in the hydration properties of the monosulfonated and disulfonated polysulfones only becomes pronounced at higher ion contents. In terms of the ion content, mBPS75 is between dBPS33 (at 66\%) and dBPS40 (at 80\%) used in the DSC measurements. However, the amount of ``crystallizable'' water computed with MD simulations using the ``bulk-like'' idea is somewhat higher than the amount of melting water experimentally measured for both dBPS33 and dBPS40, as shown in Fig.~\ref{fig:compmelting}(b). This difference is likely due to the criterion used to identify ``crystallizable'' water in the simulations, where water molecules are classified as ``bulk-like'' and therefore ``crystallizable'' as long as they are close enough to form four hydrogen bonds with surrounding water molecules, no matter if they have the rotational freedom to form a crystalline lattice as in ice. A better agreement between the simulation and experimental results may be achieved if a more sophisticated criterion for crystallizability is adopted.

\par
\begin{figure}[htb]
    \centering
    \includegraphics[width=0.5\textwidth]{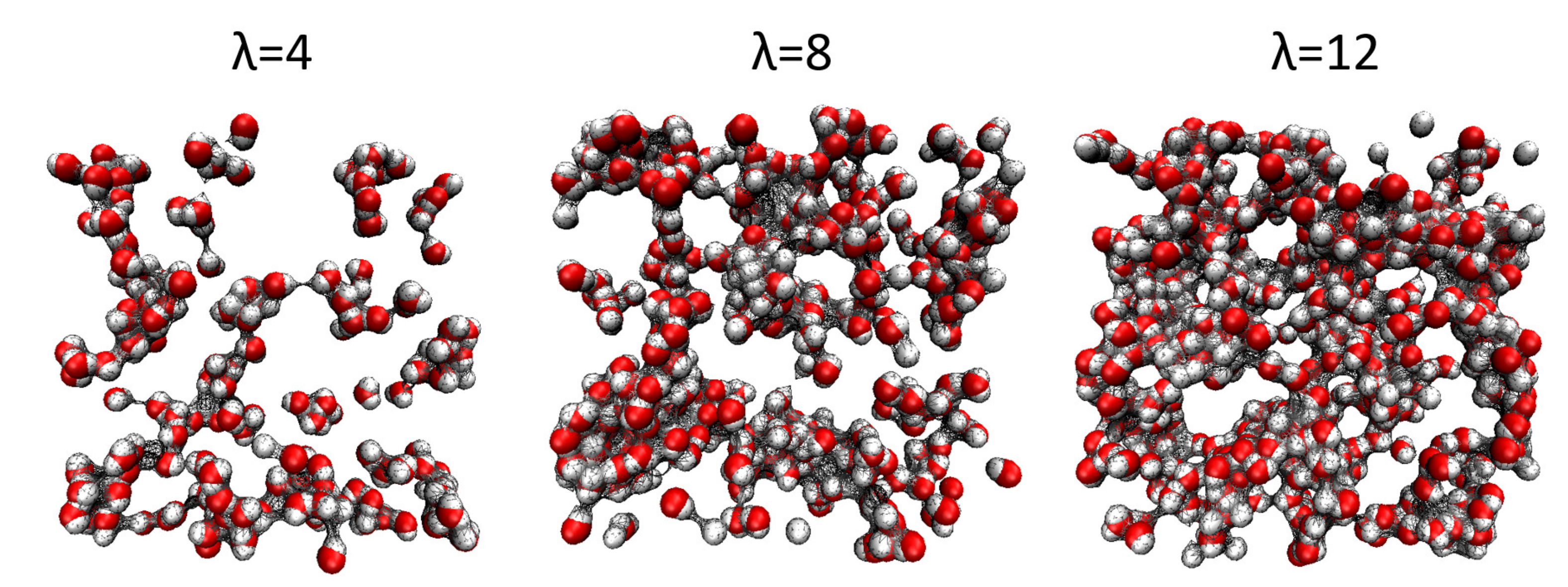}
    \caption{Snapshots of the dBPS25 system at $\lambda=$4, 8, and 12 from MD simulations showing the enhanced formation of water clusters as $\lambda$ is increased. For clarity, a 20~\AA~ slice is shown in each snapshot and only water molecules are visualized. Grey webs are included as a guide to the eye.}
    \label{fig:aggregates}
\end{figure}

 The amount of crystallizable water presented in Fig.~\ref{fig:compmelting}(b) indicates that as the water uptake increases for a given sulfonated polysulfone, the degree of clustering, and therefore the likelihood of a water molecule being bulk-like, is enhanced. This trend is visually demonstrated with three snapshots from MD simulations shown in Fig.~\ref{fig:aggregates}, which are for the dBPS25 system at $\lambda=$4, 8, and 12. As more water is added to the system, the sizes of water clusters clearly become larger. As a result, the amount of bulk-like water that can form four hydrogen bonds, a criterion used to identify ``crystallizable'' water in the simulations, increases with an increasing $\lambda$. Figure~\ref{fig:aggregates} also shows that when the polymer absorbs more water, initially separated water clusters start to merge and the number of water clusters decreases.

\par
\begin{figure}[htb]
    \centering
    \includegraphics[width=0.45\textwidth]{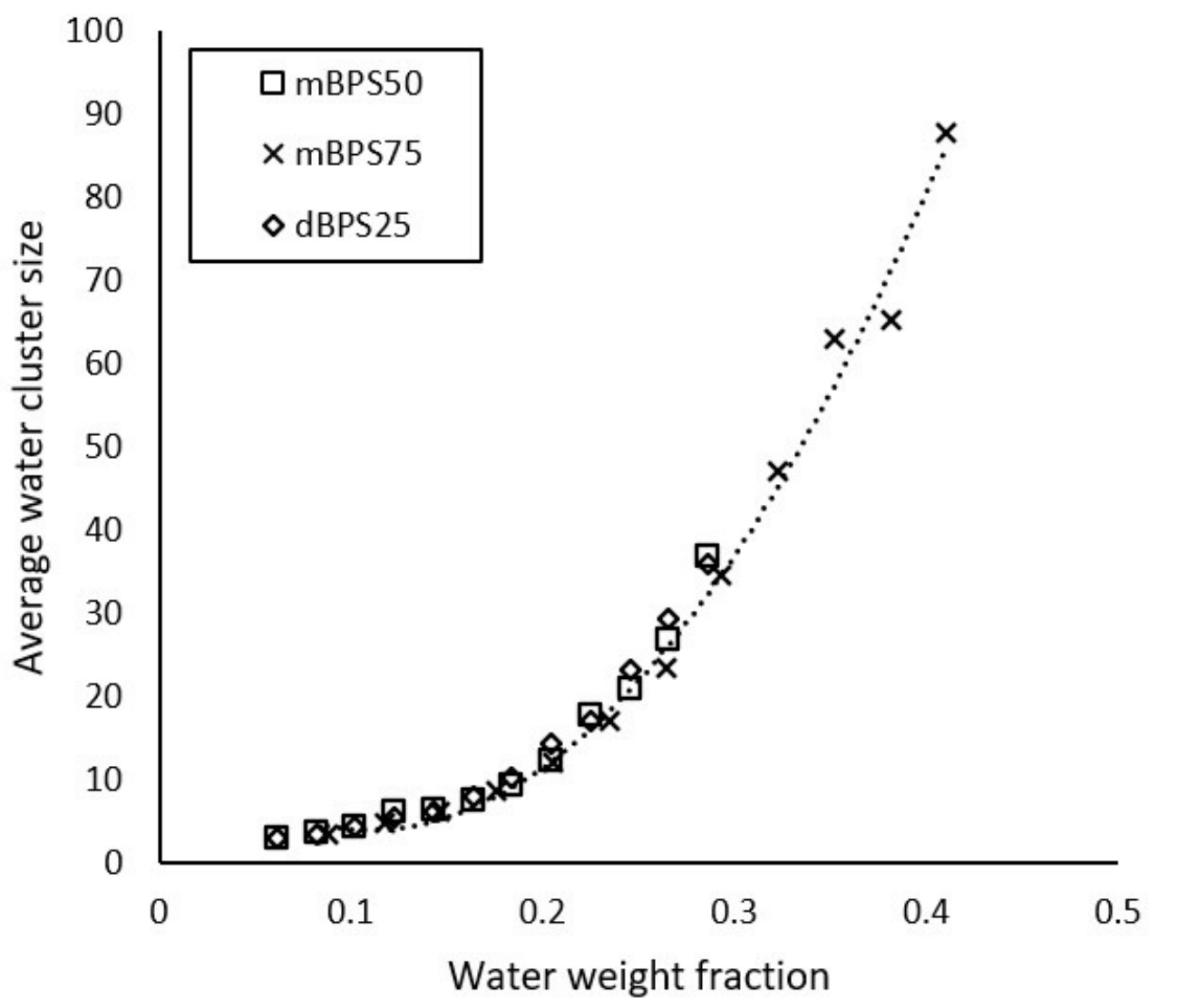}
    \caption{Average water cluster size vs. water weight fraction from MD simulations for mBPS50 (squares), mBPS75 (multiplication signs), and dBPS25 (diamonds). The dotted line is a second order polynomial fit to the data for mBPS75.}
    \label{fig:clustersize}
\end{figure}

\begin{figure}[htb]
    \centering
    \includegraphics[width=0.45\textwidth]{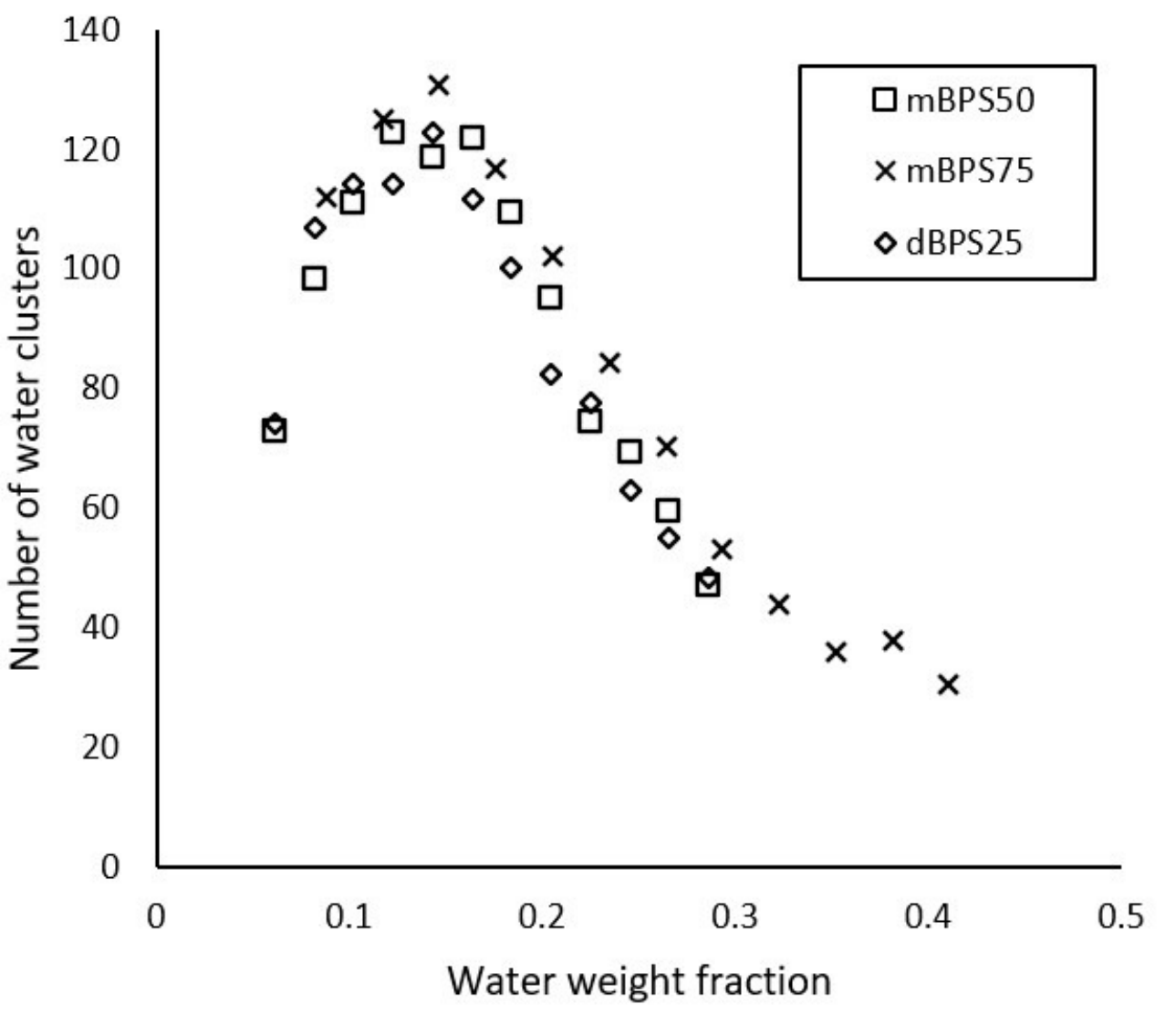}
    \caption{Average number of water clusters vs. water weight fraction from MD simulations for mBPS50 (squares), mBPS75 (multiplication signs), and dBPS25 (diamonds).}
    \label{fig:clusternum}
\end{figure}

 To quantify the development of water clusters visualized in Fig.~\ref{fig:aggregates}, we perform a cluster analysis for the absorbed water, similar to the distance-based clustering analysis conducted by Abbott and Frischknecht.\cite{abbott_nanoscale_2017} Two water molecules are considered clustered when their oxygen atoms are within a cutoff separation of 3.4~\AA. The results on the average cluster size and the number of clusters are included in Figs.~\ref{fig:clustersize} and \ref{fig:clusternum}, respectively. To compare polymers with different ion contents, we plot these quantities as a function of the weight fraction of water. All polysulfones modeled show a similar behavior of water clustering. As more water is absorbed by a polymer, the average size of water clusters grows almost quadratically with respect to the water's weight fraction. The number of water clusters initially increases with its weight fraction, peaks at a weight fraction about 15\%, and then decreases as more water is loaded into the polymer. At low water contents, water starts to fill the free volume in the polymer and more clusters are formed. In this process, the volume of the polymer does not change significantly. However, when the amount of water in the polymer exceeds some threshold, water clusters start to merge and their number decreases, consistent with the visualizations in Fig.~\ref{fig:aggregates}. At the same time, the polymer starts to swell at high water contents. The maximum in Fig.~\ref{fig:clusternum} represents the transition from the free-volume filling regime to the swelling regime, which is also reflected by the measured density of the hydrated polymers included in Table \ref{TB1}.

\section{Discussion}

\par According to the three-state model outlined in Sec.~\ref{Intro}, whether or not water molecules within an ionic polymer can melt is related to how they are associated with the ions and polar groups. However, our results indicate that melting water in an ionic polymer is more strongly correlated with water molecules in the right aggregation state. The comparison between the simulation and experimental results indicates that melting water consists of water molecules that have a bulk-like hydrogen-bonding environment around them and are thus crystallizable. Melting refers to the transition of water (or water clusters inside a hydrated polymer) from a crystalline phase to a liquid phase. That is, melting water is crystallizable and vice versa. In order to crystallize, water molecules must have the ability to hydrogen bond sufficiently to neighboring water molecules. This criterion of identifying melting water on the basis of a bulk-like hydrogen-bonded network is further supported by the FTIR data. These data indicate that the amount of melting water is associated, not with the ``free'' OH bond stretching signature, but with the most constrained and completely hydrogen-bonded OH stretching region at lower wavenumbers. Based on this evidence, rather than ``unbound'' and ``bound'', the states of water in an ionic polymer are more accurately termed crystallizable or ``aggregated'' for water molecules that show a melting behavior, and non-crystallizable or ``isolated'' for those that do not. The term ``bulk-like'' to describe crystallizable or non-confined water molecules is already in common use in some fields \cite{roget_bulk-like_2019} and also appears to be appropriate for the hydrated ionic polymers studied here.

\par Although the above discussion can be viewed as semantics, the way that we think about the nature of melting water in ionic polymers with aromatic backbones has far-reaching implications for designing polymers for membrane applications.\cite{daryaei2017-1,choudhury2019} If nonmelting water is thought of as being bound to or immobilized by ions through hydration, then the main parameter that we could change to control the amount of melting water in a hydrated polymer would be the ion content of the polymer. If that were the case, the amount of melting water should be very similar among polymers with the same amount of ions, regardless of their other characteristics (e.g. the placement of the ionic groups along the polymer backbone). We have shown here that this is not the case and a small change in the inter-ion spacing along the backbone of a polymer with a fixed ion content can cause a measurable difference in the amount of melting water in the polymer. Therefore, thinking of nonmelting water as ``bound water'' is not a useful guide to the design of ionic polymers, especially those with aromatic backbones, for applications where the amount of melting water is an important design parameter.

\par In contrast, if we think of melting water as water molecules that are clustered together within a hydrated polymer, then the understanding of the experimental results is straightforward, and the design of ionic polymers for applications involving ion hydration and water transport becomes more efficient. Our MD simulations suggest that up to 75\% of water molecules in the sulfonated polysulfones are associated with the ions. Therefore, placing ions close together on the backbone can cause the associated water molecules to merge and form aggregates that possess certain properties of bulk water, even if the ion content and water content of the polymer are low. Conversely, if ions are well distributed along the backbone, then their associated water molecules remain isolated from each other and water aggregates will not form until the ion content is made sufficiently high.

\par One aspect that remains unclear is the nature of ``loosely bound'' water. Previous DSC measurements of ionic polymers with low glass transition temperatures have found two distinct melting endotherms.\cite{roy2006} The lower temperature endotherm is usually attributed to ``loosely bound" water molecules. In the sulfonated polysulfones investigated here, the peak location in the water-melting endotherm varies from 0 \degree C for high-water-content samples to -16 \degree C for low-water-content samples but there remains a single melting endotherm for all samples. A similar behavior has been observed for other ionic polymers with aromatic backbones and high glass transition temperatures.\cite{wu2010} Terahertz spectroscopy measurements of aqueous solutions of biomolecules and organic molecules have pointed to the existence of water molecules outside the primary hydration shells of the solute molecules but distinct from bulk water with respect to dynamics and hydrogen bonding.\cite{Charkhesht2018, Charkhesht2019} However, more research is needed to understand the connection between the results on water dynamics revealed by terahertz spectroscopy and those from DSC measurements. Some authors have also suggested that the existence of two melting endotherms is related to a polymer's possible glass transition and the associated impact on the nucleation rate of ice crystals.\cite{kocherbitov2016, muller-plathe_1998, loozen_anomalous_2006} This idea seems to be useful to the understanding of our experimental results, although more work on a wider range of polymers with various backbones will be needed to establish a more complete physical picture.

\section{Conclusions}

\par We combine DSC analyses, FTIR measurements, and molecular dynamics simulations to elucidate the nature of freezing/melting water in sulfonated polysulfones over a range of ion contents and water uptakes. The DSC curves of these polymer, each at its equilibrium water uptake, indicate that melting water begins to appear at a lower ion content for the disulfonated polymers than for the monosulfonated polymers. Similarly, analysis of the FTIR OH stretching band indicates that in the disulfonated polymers there is a strong transition in the bulk-like OH vibrations when the ion content exceeds a certain threshold. In contrast, the subpeak area grows gradually with an increasing ion content for the monosulfonated polymers and no sharp transition is detected in the range of ion contents investigated here. Both DSC and FTIR results thus reveal that at a fixed ion content, the spacing of the ions along the backbone significantly impacts the amount of bulk-like water within a hydrated ionic polymer.

\par We have tested two hypotheses for the physical origin of melting water within hydrated ionic polymers. In one hypothesis, the ``nonprimary'' water molecules outside the primary hydration shells of ions and polar groups are regarded as melting water. In the other hypothesis, water molecules that can hydrogen bond to at least four neighboring water molecules, i.e., ``bulk-like'', and therefore can melt and crystallize are classified as melting water. The comparison between our experimental and simulation results supports the latter hypothesis based on crystallizable water.

\par The key point is that there is a distinction between the ion-association state and the hydrogen-bonding state of water molecules in a hydrated ionic polymer. A water molecule not strongly associated with an ion (i.e., not within the primary hydration shell of any ion) may not have a sufficient number of adjacent water molecules to establish a bulk-like hydrogen-bonded network and such water molecules still do not contribute to the melting process. The distinction has important implications for the design of ionic polymers for various applications because it shows that the amount of aggregated or bulk-like water can be somewhat decoupled from the ion content of a polymer. Instead, the amount of bulk-like water in a hydrated polymer is directly related to the spatial distribution of ions in the polymer. Since the presence of bulk-like water may have a negative impact on the selectivity of a membrane, the ions should be distributed in such a way that the amount of bulk-like water is minimized for membrane applications. Our results indicate that this goal can be achieved by spacing the ions in the polymer as evenly as possible via controlled synthesis.

\section*{Acknowledgments}
The authors acknowledge the Donors of the American Chemical Society Petroleum Research Fund (PRF \#56103-DNI6), for support of this research (C.W. and S.C.). This research was also supported by a 4-VA Collaborative Research Grant (S.C.) and the Virginia Tech College of Engineering (B.V. and J.L.). The authors acknowledge Advanced Research Computing at Virginia Tech (URL: http://www.arc.vt.edu) for providing computational resources and technical support that have contributed to the results reported within this paper. S.C. also gratefully acknowledges the support of NVIDIA Corporation with the donation of the Tesla K40 GPUs used for this research.

%\bibliographystyle{ieeetr}
%\bibliography{Melting}

\begin{thebibliography}{10}

\bibitem{Harrison2005}
W.~L. Harrison, M.~A. Hickner, Y.~S. Kim, and J.~E. McGrath,
\newblock Fuel Cells {\bf 5}, 201 (2005).

\bibitem{Tang2014}
Y.~Tang et~al.,
\newblock Polym. Chem. {\bf 5}, 6097 (2014).

\bibitem{daryaei2017-1}
A.~Daryaei et~al.,
\newblock ACS Appl. Mater. Interfaces {\bf 9}, 20067 (2017).

\bibitem{daryaei2017-2}
A.~Daryaei et~al.,
\newblock Polymer {\bf 132}, 286 (2017).

\bibitem{choudhury2019}
S.~R. Choudhury et~al.,
\newblock Polymer  (2019).

\bibitem{tran2019}
T.~Tran, C.~Lin, S.~Chaurasia, and H.~Lin,
\newblock J. Membr. Sci. {\bf 574}, 299 (2019).

\bibitem{woo_2003}
Y.~Woo, S.~Y. Oh, Y.~S. Kang, and B.~Jung,
\newblock J. Membr. Sci. {\bf 220}, 31 (2003).

\bibitem{kocherbitov2010}
V.~Kocherbitov, S.~Ulvenlund, L.-E. Briggner, M.~Kober, and T.~Arnebrant,
\newblock Carbohydr. Polym. {\bf 82}, 284 (2010).

\bibitem{karlsson2002}
L.~E. Karlsson, B.~Wesslén, and P.~Jannasch,
\newblock Electrochim. Acta {\bf 47}, 3269 (2002).

\bibitem{wu2010}
X.~Wu, G.~He, S.~Gu, Z.~Hu, and X.~Yan,
\newblock Chem. Eng. J. {\bf 156}, 578 (2010).

\bibitem{luck1980}
W.~A.~P. Luck,
\newblock The structure of aqueous systems and the influence of electrolytes,
\newblock in {\em Water in {Polymers}}, edited by S.~P. Rowland, volume 127,
  pages 43--71, American Chemical Society, Washington, D. C., 1980.

\bibitem{ping2001}
Z.~H. Ping, Q.~T. Nguyen, S.~M. Chen, J.~Q. Zhou, and Y.~D. Ding,
\newblock Polymer {\bf 42}, 8461 (2001).

\bibitem{roy2017}
A.~Roy et~al.,
\newblock Polymer {\bf 111}, 297 (2017).

\bibitem{zhang_molecular_2017}
Q.~Zhang, T.~Wu, C.~Chen, S.~Mukamel, and W.~Zhuang,
\newblock Proc. Natl. Acad. Sci. USA {\bf 114}, 10023 (2017).

\bibitem{braun_rotational_2016}
D.~Braun, M.~Schmollngruber, and O.~Steinhauser,
\newblock Phys. Chem. Chem. Phys. {\bf 18}, 24620 (2016).

\bibitem{marque2010}
G.~Marque, J.~Verdu, V.~Prunier, and D.~Brown,
\newblock J. Polym. Sci. B: Polym. Phys. {\bf 48}, 2312 (2010).

\bibitem{dogonadze_chemical_1985}
R.~R. Dogonadze,
\newblock {\em The Chemical Physics of Solvation},
\newblock Number 38A in Studies in Physical and Theoretical Chemistry,
  Elsevier, Amsterdam; New York, 1985.

\bibitem{Charkhesht2018}
A.~Charkhesht, C.~K. Regmi, K.~R. Mitchell-Koch, S.~Cheng, and N.~Q. Vinh,
\newblock J. Phys. Chem. B {\bf 122}, 6341 (2018).

\bibitem{Charkhesht2019}
A.~Charkhesht et~al.,
\newblock J. Phys. Chem. B {\bf 123}, 8791 (2019).

\bibitem{pastorczak2014}
M.~Pastorczak et~al.,
\newblock Colloid Polym. Sci. {\bf 292}, 1775 (2014).

\bibitem{kocherbitov2016}
V.~Kocherbitov,
\newblock Carbohydr. Polym. {\bf 150}, 353 (2016).

\bibitem{thouvenin_2002}
M.~Thouvenin, I.~Linossier, O.~Sire, J.-J. Péron, and K.~Vallée-Réhel,
\newblock Macromolecules {\bf 35}, 489 (2002).

\bibitem{lasagabaster_2006}
A.~Lasagabaster, M.~J. Abad, L.~Barral, and A.~Ares,
\newblock Eur. Polym. J. {\bf 42}, 3121 (2006).

\bibitem{smedley_measuring_2015}
S.~B. Smedley, Y.~Chang, C.~Bae, and M.~A. Hickner,
\newblock Solid State Ionics {\bf 275}, 66 (2015).

\bibitem{ben-amotz_2019}
D.~Ben-Amotz,
\newblock J. Am. Chem. Soc. {\bf 141}, 10569 (2019).

\bibitem{verma_2018}
P.~K. Verma et~al.,
\newblock J. Phys. Chem. B {\bf 122}, 2587 (2018).

\bibitem{tainter_2013}
C.~J. Tainter, Y.~Ni, L.~Shi, and J.~L. Skinner,
\newblock J. Phys. Chem. Lett. {\bf 4}, 12 (2013).

\bibitem{luo_engineering_2019}
H.~Luo, K.~Chang, K.~Bahati, and G.~M. Geise,
\newblock Environ. Sci. Technol. Lett. {\bf 6}, 462 (2019).

\bibitem{luo_functional_2019}
H.~Luo, K.~Chang, K.~Bahati, and G.~M. Geise,
\newblock J. Membr. Sci. {\bf 590}, 117295 (2019).

\bibitem{skinner_2008_2}
J.~L. Skinner, B.~M. Auer, and Y.-S. Lin,
\newblock Vibrational line shapes, spectral diffusion, and hydrogen bonding in
  liquid water,
\newblock in {\em Advances in Chemical Physics}, edited by S.~A. Rice, pages
  59--103, John Wiley \& Sons, Ltd, 2009.

\bibitem{venkatnathan2007}
A.~Venkatnathan, R.~Devanathan, and M.~Dupuis,
\newblock J. Phys. Chem. B {\bf 111}, 7234 (2007).

\bibitem{vishnyakov2008}
A.~Vishnyakov and A.~V. Neimark,
\newblock J. Phys. Chem. B {\bf 112}, 14905 (2008).

\bibitem{pozuelo2006}
J.~Pozuelo, E.~Riande, E.~Saiz, and V.~Compa{\~n},
\newblock Macromolecules {\bf 39}, 8862 (2006).

\bibitem{merinov2013}
B.~V. Merinov and W.~A. Goddard~III,
\newblock J. Membr. Sci. {\bf 431}, 79 (2013).

\bibitem{huo2019}
J.~Huo et~al.,
\newblock Int. J. Hydrog. Energy {\bf 44}, 3760 (2019).

\bibitem{bahlakeh2013}
G.~Bahlakeh, M.~Nikazar, and M.~Mahdi Hasani-Sadrabadi,
\newblock J. Membr. Sci. {\bf 429}, 384 (2013).

\bibitem{BAHLAKEH2012}
G.~Bahlakeh, M.~Nikazar, M.-J. Hafezi, E.~Dashtimoghadam, and M.~M.
  Hasani-Sadrabadi,
\newblock Int. J. Hydrog. Energy {\bf 37}, 10256  (2012).

\bibitem{tripathy_molecular_2017}
M.~Tripathy, P.~B.~S. Kumar, and A.~P. Deshpande,
\newblock J. Phys. Chem. B {\bf 121}, 4873 (2017).

\bibitem{roy2006}
A.~Roy et~al.,
\newblock J. Polym. Sci. B: Polym. Phys. {\bf 44}, 2226 (2006).

\bibitem{daryaei2017}
A.~Daryaei,
\newblock {\em Synthesis and {Characterization} of {Linear} and {Crosslinked}
  {Sulfonated} {Poly}(arylene ether sulfone)s: {Hydrocarbon}-based {Copolymers}
  as {Ion} {Conductive} {Membranes} for {Electrochemical} {Systems}},
\newblock Doctoral dissertations, Virginia Polytechnic Institute and State
  University, Blacksburg, VA, 2017.

\bibitem{muller-plathe_1998}
F.~M{\"u}ller-Plathe,
\newblock Macromolecules {\bf 31}, 6721 (1998).

\bibitem{sundell2014}
B.~J. Sundell,
\newblock {\em Synthesis and {Characterization} of {Poly}(arylene ether
  sulfone)s for {Reverse} {Osmosis} {Water} {Purification} and {Gas}
  {Separation} {Membranes}},
\newblock Doctoral dissertation, Virginia Polytechnic Institute and State
  University, Blacksburg, VA, 2014.

\bibitem{vondrasek2021}
B.~Vondrasek, C.~Wen, S.~Cheng, J.~S. Riffle, and J.~J. Lesko,
\newblock Macromolecules {\bf 54}, 302 (2021).

\bibitem{sloughnodate}
C.~G. Slough,
\newblock Cleaning of {DSC} pans,
\newblock Accessed February 10, 2021.

\bibitem{Ruscic_2013}
B.~Ruscic,
\newblock J. Phys. Chem. A {\bf 117}, 11940 (2013).

\bibitem{auer_ir_2008}
B.~M. Auer and J.~L. Skinner,
\newblock J. Chem. Phys. {\bf 128}, 224511 (2008).

\bibitem{yangnodate}
P.~Yang and P.~T. Mather,
\newblock Thermal analysis to determine various forms of water present in
  hydrogels,
\newblock Accessed March 10, 2021.

\bibitem{abbott_nanoscale_2017}
L.~J. Abbott and A.~L. Frischknecht,
\newblock Macromolecules {\bf 50}, 1184 (2017).

\bibitem{roget_bulk-like_2019}
S.~A. Roget, P.~L. Kramer, J.~E. Thomaz, and M.~D. Fayer,
\newblock J. Phys. Chem. B {\bf 123}, 9408 (2019).

\bibitem{loozen_anomalous_2006}
E.~Loozen, K.~Durme, E.~Nies, B.~Mele, and H.~Berghmans,
\newblock Polymer {\bf 47}, 7034 (2006).

\end{thebibliography}

\end{document}